\documentclass[prd,tightenlines,nofootinbib,superscriptaddress,twocolumn]{revtex4}
\usepackage[colorlinks=true,linkcolor=blue,citecolor=purple,linktocpage=true]{hyperref}

\usepackage{mathtools}
\usepackage{amssymb}
\usepackage{dsfont}
\usepackage{lipsum}
\usepackage{mathtools}
\usepackage{xcolor}

\usepackage[makeroom]{cancel}

\def\be{\begin{equation}}
\def\ee{\end{equation}}
\def\bea{\begin{eqnarray}}
\def\eea{\end{eqnarray}}
\def\beq{\begin{eqnarray}}
\def\eeq{\end{eqnarray}}
\def\nn{\nonumber}
\newcommand{\Ff}{{\mathstrut_1 F_1}}
\newcommand{\ii}{{\rm i}}
\definecolor{dblue}{rgb}{0, 0.25, 1}
\def\nn{\nonumber}
\def\dd{\mathrm{d}}

\begin{document}

\title{Rutherford scattering of quantum and classical fields}

\author{{\bf Martin Pijnenburg}}\email{martin.pijnenburg@unige.ch}
\affiliation{Département de Physique Théorique and Center for Astroparticle Physics, Université de Genève, Quai E. Ansermet 24, CH-1211 Genève 4, Switzerland}
\author{{\bf Giulia Cusin}}\email{giulia.cusin@iap.fr}
\affiliation{Institut d'Astrophysique de Paris, UMR-7095 du CNRS, Paris, France}
\affiliation{Département de Physique Théorique and Center for Astroparticle Physics, Université de Genève, Quai E. Ansermet 24, CH-1211 Genève 4, Switzerland}
\author{{\bf Cyril Pitrou}}\email{pitrou@iap.fr}
\affiliation{Institut d'Astrophysique de Paris, UMR-7095 du CNRS, Paris, France}
\author{{\bf Jean-Philippe Uzan}}\email{uzan@iap.fr}
\affiliation{Institut d'Astrophysique de Paris, UMR-7095 du CNRS, Paris, France}

\date{\today}
\begin{abstract}
Quantum Rutherford scattering and scattering of classical waves off Coulomb-like potentials have similar formal structures and can be studied using the same mathematical techniques. In both contexts, the long-range nature of the interaction leads to a divergent total cross-section, which has been interpreted and regularized in various ways in the past literature. We review in detail the origin of this divergence, in both real and multipole spaces, and show that it arises from incorrectly using approximations out of their domain of validity. We also stress that although classical and quantum Rutherford scattering share the same formalism, the natures of the associated physical observables differ. We comment on the role of interference: while interference can be safely neglected in a quantum context (due to the fact that the observable quantity is a flux, and the incoming flux is collimated), in a classical context one expects to see a superposition of transmitted and scattered waves in a broad region downstream of the target, hence a cross-section is not connected to any physically observable quantity.
\end{abstract}
\maketitle




\section{Introduction} 

Scattering phenomena are ubiquitous in physics, they range from classical mechanics (e.g. massive particles in a potential), quantum mechanics (e.g. scattering in the Schr\"odinger framework) and relativity (e.g. light or gravitational waves scattering off astrophysical sources). In the latter case, the scattering of classical fields of various spins off massive objects is mediated by the metric describing a given curved background (typically Schwarzschild or Kerr), and described in the framework of general relativity, see e.g. Refs.~\cite{Schiff, futterman_handler_matzner_1988} for a review. 

The general structure of the Schr\"odinger equation in $D=3$ spatial dimensions for a monochromatic wave of frequency $\omega$ and energy $E=\hbar\omega$ (associated to a massive particle) scattering off a potential $V$ is
\be\label{e.schro}
\left(–\frac{\hbar^2}{2m}\Delta + V\right)\psi = E\psi\,,
\ee
where $\Delta$ is the Laplacian operator. The term on the RHS arises from the time derivative $\ii\hbar\partial_t\psi(r,t)$ acting on a monochromatic wave of frequency $\omega$, i.e. $\psi(r,t)\propto \psi(r)\exp\left(-\ii \omega t \right)$.

In most cases one does not identify or need an exact solution of the full scattering equation but rather solve it in a ``large distance" limit in which the solution behaves as the superposition of a spherical (scattered) wave and incident plane wave. Indeed, the definition of the differential cross-section $\dd\sigma/\dd\Omega$ relies on the assumption that it is possible to split the wave solution of Eq.~(\ref{e.schro}) into the sum of incoming and outgoing waves as
\be\label{split}
\psi=\psi_{\text{in}}+\psi_{\text{scat}}\,.
\ee
Then, one defines the differential cross-section in a direction $\theta$ as the flux of the current of the scattered waves in a surface $\dd S$  normalized to the initial flux, i.e.
\be
\label{IntroToCrossSection}
{\bf J}_{\rm scat} \cdot \dd S {\bf e}_r \equiv \left|{\bf J}_{\rm in}\right|\dd\sigma(\theta)\,, 
\ee
where $\dd S=r^2\dd\Omega$, and $r$ is a radial distance. 

In many physical situations the potential in Eq.~(\ref{e.schro}) has
poles (e.g. in $r=0$ in the case of the Coulomb or the Newtonian potentials). As a consequence, the
first derivative of the solution cannot be continuous at the pole
location but the solution may nonetheless be well-defined and
continuous. Also, when the potential does not fall off rapidly enough, as
is the case for the Coulomb potential on which we focus hereafter, the cross-section diverges in the forward direction. Exact solutions for such
Rutherford scattering \cite{Rutherford1911} problems do actually exist and have been known for a long
time \cite{Gordon, Schiff, Matzner,
  futterman_handler_matzner_1988, PhysRevD.14.317, Dolan:2007ut,
  Dolan:2008kf, Gaspard} (see also Ref.~\cite{Barton:1983un} for the Rutherford
solution in two dimensions). Since these exact solutions are
well-defined everywhere in space, they allow one to track back the origin
of the divergence of approximate solutions. Nevertheless, in the
literature several \emph{(un)physical} explanations are presented to
justify the presence of these divergences, which we will review in this work.  
This article provides a pedagogical example stressing the importance of
correctly stating the regime of validity of an approximate solution in
order to correctly draw a physical interpretation from it. We also show that
great care has to be taken in the inversion of limits, to avoid
misinterpreting anomalous behaviours of results which are in fact
simply a consequence of an incorrect extrapolation of the approximate
solution itself. There is actually no need to invoke methods to regularise Rutherford-like $\theta\to0$ divergences nor to look for a physical explanation to justify its presence. 

This paper is structured as follows. 
First, in section \ref{s.ReviewRutherford} we briefly review standard
concepts in scattering problems, and we apply the usual procedure to
compute the scattering cross-section of Rutherford
scattering. We show that the amplitude of the resulting scattered wave is divergent in the forward direction (hence the cross-section has the same type of divergence).
In section \ref{SectionExactToRutherford} we show that the  Rutherford
scattering equation possesses an exact solution, and we discuss its behaviour. In particular, the exact solution is
well-defined (not divergent) everywhere in space. We then identify the
parameter allowing to split the total solution in terms of
incoming and outgoing waves, see Eq.\,(\ref{split}). The
``large-$r$'' limit appears as an approximation valid only for large
values of this dimensionless parameter which is a combination of $r$ and
$\theta$. As a consequence, for any given radial distance $r$, one can
define a region around the forward direction inside which the
expansion (\ref{split}) is not valid. Hence, the Rutherford formula
for the scattering cross-section only holds outside this
region. The $\theta=0$ divergence arises from the
extrapolation of the approximate solution in a regime where it never
holds. The exact solution of Rutherford scattering can also be obtained using a multipolar
decomposition, and this is detailed in section \ref{s.ellspace} where we show the connections between the approximate solutions in real and
multipolar spaces. Finally, section \ref{class} turns to a different
problem: classical scattering of waves off Coulomb-like
potentials. Mathematically the equation governing this scattering
problem is very similar to the Rutherford equation describing quantum
scattering~\cite{Rutherford1911}. However, we stress some
caveats one has to be aware of when relying on this formal analogy. In
particular, we show that a definition of a classical cross-section in
this context does not correspond to any observable quantity and can be
misleading when trying to interpret physical effects in this classical
framework.

\section{Review of Rutherford}
\label{s.ReviewRutherford}

The starting point of our scattering problem is the Schr\"odinger equation \eqref{e.schro} where the solution $\psi$ contains an incident plane wave, incoming along the $e_z$-axis 
\be \label{e.planew}
\psi_{\rm in}=\hbox{e}^{\ii kz}\,. 
\ee
We use spherical coordinates $(r,\theta,\varphi)$ so that
$z=r\cos\theta$  and ${\bf e}_z = \cos\theta {\bf e}_r -
\sin\theta{\bf e}_\theta$. The spherically symmetric Coulomb potential for the Rutherford
problem is of the form
\be
\label{CoulombPotential}
V=\frac{A}{r}\,,
\ee
where $A$ is a constant. Defining
\be
\hbar k=\sqrt{2mE},\quad
\gamma = \frac{mA}{\hbar^2k}\,,\quad \rho = k r\,,
\ee 
the Schr\"odinger equation~(\ref{e.schro}) reads
\be\label{e.schro2}
\left(\Delta + k^2  - 2\frac{\gamma k}{r}\right)\psi = 0\,,
\ee
or equivalently
\be\label{e.schro2b}
\left(\partial_\rho^2 +\frac{2}{\rho}\partial_\rho + \frac{1}{\rho^2}\Delta_2 + 1  - 2\frac{\gamma}{\rho}\right)\psi = 0\,,
\ee
where the Laplacian on the unit sphere is
\be
\Delta_2 \psi = \frac{1}{\sin \theta} \frac{\partial}{\partial \theta}\left(\sin \theta \frac{\partial \psi}{\partial \theta}\right) + \frac{1}{\sin^2 \theta} \frac{\partial^2 \psi}{\partial \varphi^2}\,.
\ee

In the standard analysis, one considers the problem in a regime where the waveform can be split into an incoming plane wave and an outgoing spherical wave as in Eq.~(\ref{split}). This regime is commonly referred to as the $r\rightarrow+\infty$ limit (however, as we will show in the next section, this definition induces misconceptions when considering the infinitesimal scattering angle limit).  

In the standard scattering literature, the incoming wave~(\ref{e.planew}) is superposed to the scattered wave, parameterised as 
\be
\label{e.SphericalScatteredWave}
\psi_{\rm scat} = f(\theta,\varphi)\frac{\hbox{e}^{\ii kr}}{r}\,.
\ee
The scattering amplitude $f$ is obtained by solving (\ref{e.schro}), with proper boundary conditions. 
Since the potential \eqref{CoulombPotential} is spherically symmetric,
and having chosen that the incoming wave travels along the azimuthal direction
$z$, it is clear by symmetry that $f$ depends only on the angle $\theta$, i.e. $f=f(\theta)$. 

So far, the mathematical description of Rutherford scattering relies on an (infinitely extended) plane wave, scattering off a Coulomb potential. The fundamental object of this description is a wavefunction, which is related to a probability density and a current, which are physical observables.
In general, the current associated to a given waveform $\psi$ is given by
\be
{\bf J}[\psi] = \frac{\hbar }{m}{\rm Im}\left[\psi^*\mathbf{\nabla}\psi\right]\,,
\label{J}
\ee
such that the outgoing current (use $\psi = \psi_\text{scat}$ in \eqref{J}) is
\be\label{Jscat}
{\bf J}_{\rm scat} = \frac{\hbar k}{m}\frac{\vert f\vert^2}{r^2}{\bf e}_r + {\cal O}(1/r^3)\,.
\ee
In a similar fashion, we can compute $\left|{\bf J}_{\rm in}\right| = \frac{\hbar k}{m}$.
From Eq.\,\eqref{IntroToCrossSection}, the differential cross-section
is related to $f$ by
\be\label{sigmadef}
\frac{\dd \sigma}{\dd\Omega}(\theta) \equiv\frac{{\bf J}_{\rm scat}.{\bf e}_r \rho^2}{k^2 \left|{\bf J}_{\rm in}\right|}=\vert f(\theta) \vert^2\,.
\ee
It is important to remind that one has ({\it i}) taken the large-$r$ limit
on the wave solution and ({\it ii}) extracted the scattered wave from the
total wave before computing the current. Indeed ${\bf J}[\psi]
-{\bf J}[\psi_{\rm in}]\not={\bf J}[\psi_{\rm scat}]$, where the difference
arises from the presence of interference terms.  However, from an experimental point of view, one sends a collimated beam of particles against a target, and we measure a (steady) flux of scattered particles. We need to adapt the mathematical description to this experimental setting, and as a consequence we can ignore interference, see discussion at the end of section \ref{Sect:Inter}.

There are different methods to compute $f(\theta)$ and the scattered
wave in the so-called large-$r$ limit. We provide here the result based on the Born approximation. We note  ${\bf k}$ the initial wavevector and ${\bf k'}$ the wavevector of the
scattered wave. The scattered wave is spherical
by assumption, therefore $\bf k' = {\bf |k'|}{\bf e}_r$. We assume that
the scattering potential does not absorb momentum (elastic collision,
fixed target) so that ${\bf |k|=|k'|} =k$, i.e. ${\bf k'}=k {\bf e}_r$. Defining $\bf q \equiv  {\bf k} -{\bf k'}$, the Born
approximation yields (see Refs.~\cite{Bluegel, TaylorBook, Tong})
\bea\label{Born1f(theta)}
f(\theta) &=& -\frac{2 m}{4 \pi \hbar^2} \int {\rm e}^{-\ii {\bf q}\cdot {\bf r}} V(r)\dd^3 {\bf r}\\
&=&-\frac{2 m}{\hbar^2}\frac{1}{q}\int_0^\infty V(r) r \sin(qr) \dd r \,,\nonumber
\eea
where $q\equiv|{\bf q}|=2k\sin(\theta/2)$ and $\theta$ is the
deflection angle between ${\bf k}$ and ${\bf k'}$, namely between
${\bf e}_z$ and ${\bf e}_r$, which is simply the $\theta$ angle of spherical coordinates.

The integral (\ref{Born1f(theta)}) is computed using that for any
power law potential in $D$ dimensions we have
\be
\int {\rm e}^{-\ii {\bf q}\cdot {\bf r}} r^{-\alpha}\dd^D {\bf r} = \pi^{\alpha - \frac{D}{2}} \frac{\Gamma(\frac{D-\alpha}{2})}{\Gamma(\alpha/2)} \left(\frac{2\pi}{q}\right)^{D-\alpha}\,,
\ee
where in our case $D=3$, and which is valid in the sense of (Fourier transform of) distributions for $0 < \alpha <D$. In particular for the Coulomb potential we obtain
\be \label{RutherfordAmplitudeCS}
f(\theta)  = - \frac{\gamma}{2 k \sin^2(\theta/2)}\,,\quad 
\frac{d\sigma}{d\Omega} = \frac{\gamma^2}{4 k^2\sin^4(\theta/2)} \,.
\ee
The differential cross-section is exactly the one found for the scattering of a massive
particle off a Coulomb potential in classical mechanics, a textbook
result known as \emph{classical Rutherford scattering cross-section},
which is usually obtained from the conservation of energy, angular momentum and Runge-Lenz vectors.

We observe that the cross-section is divergent for $\theta\to 0$.  
In the literature different explanations have been provided to justify the presence of such a divergence.  Often this divergence is claimed to be unphysical due to the fact that the Coulomb potential itself is unphysical because of screening:  a bare charge in vacuum cannot occur in Nature, and similar considerations hold when considering scattering of classical waves off a Newtonian potential. 
 Building up on this, many references argue that, to remove the
 divergence, the actual potential should be screened \cite{TaylorBook,
   MessiahQM1, TaylorCoulombScattering, Burke}, e.g. with a
 Yukawa-like suppression of the form
\be
V(r) = \frac{A}{r} \hbox{e}^{-\mu r}\,.
\label{ExponentiallyScreenedPotential}
\ee
This argument relies on modelling screening effects, and one could imagine infinitely many ways to get a smooth transition from a Coulomb potential at small scales to a zero potential at large scales.

We stress that also in the classical computation of Rutherford scattering,  the divergence is present and has been addressed in the literature. In this case, there exists a well-defined map between deflection angle and impact parameter $b$ (see e.g. Ref.~\cite{DBZ}), and considering $\theta_\text{defl}\to 0$ means considering $b\to\infty$. In other words, only particles passing infinitely far away from the potential are undeflected, and precisely those give divergent contributions to the cross-section. Thus, in the classical scattering case, the divergence issue is alleviated by arguing that in a concrete experimental setup $b\to\infty$ never happens, hence neither does $\theta_{\text{defl}}\to0$. Moreover, typically the classical situation is such that there are multiple scattering centers (e.g. in a classical gold foil model), and there is a maximum value for $b$, given by half of the inter atomic distance. This provides a  lower bound on $\theta_\text{defl}$ \cite{DBZ}. 
However, one cannot rely on this classical analogy when considering scattering of waves, as a definition of impact parameter is ambiguous in this case.

In the next section, we show that the problem of Rutherford scattering for waves is actually inherently free of divergence: the divergence is just an artifact of the fact that we take the $\theta\to0$ limit after having taken the large-distance limit (and the two limits do not commute). It follows that all the arguments mentioned above presented to justify the Rutherford divergence are rather unnecessary, if not artificial.

\section{From exact solution to the Rutherford solution}

\subsection{Exact solution}

\label{SectionExactToRutherford}
The Schr\"odinger equation~(\ref{e.schro2}) can be solved in terms of the
confluent hypergeometric function $\Ff$ as  \cite{Gordon, Schiff, Matzner,
  futterman_handler_matzner_1988, PhysRevD.14.317, Dolan:2007ut,
  Dolan:2008kf, Gaspard}
\begin{eqnarray}\label{e.solgen}
\psi=\hbox{e}^{\ii{\bf k}.{\bf r}}\hbox{e}^{-\pi\gamma/2}\Gamma(1+\ii\gamma) {}_1F_1[-\ii\gamma,1;\ii(kr - {\bf k}\cdot{\bf r})]\,.
  \end{eqnarray}
This exact solution and its normalisation have been obtained by
  \emph{imposing} that in the absence of interactions, it reduces to
  an incoming plane wave $\hbox{e}^{\ii{\bf k}.{\bf z}}$, in such a way that it matches the wished asymptotic behaviour \eqref{e.planew}. The derivation of \eqref{e.solgen} can be found in Appendix \ref{ExactSolDerivation}.

  It is convenient to introduce the notation 
\be
s\equiv 2\sin^2\frac{\theta}{2} = 1-\cos\theta\,.
\ee
The exact solution then takes the compact form
\be\label{exact}
\psi=\hbox{e}^{\ii\rho(1-s)}\hbox{e}^{-\pi\gamma/2}\Gamma(1+\ii\gamma) {}_1F_1\left[-\ii\gamma,1;\ii \rho s \right]\,.
  \ee
Note that the first exponential prefactor is nothing else but ${\rm
  e}^{\ii kz}$ since $\rho(1-s) = kz$. 
  
\subsection{Approximate solution}
  
To find the \emph{Rutherford solution}, one usually expands the function $ {}_1F_1$ for large values of its argument, i.e.  $\rho s\gg1$.  
In other words, for a fixed geometry and wave frequency ($r$ and $k$ respectively) we identify 
an angular region
$s\ll 1/(r k)$ and compute the approximate solution valid \emph{outside} this region. 

Explicitly, one makes the following expansion valid for asymptotically large values of $q$  (e.g. Refs.~\cite{AbramowitzStegun}, \cite{TemmeBook})
\begin{align}
&\Ff(a, b; q)\sim \hbox{e}^q q^{a-b}\frac{\Gamma(b)}{\Gamma(a)}\sum_{k=0}^\infty \frac{(b-a)_k (1-a)_k}{k! q^k}\nn\\
&+\hbox{e}^{\pm \ii \pi a}q^{-a}\frac{\Gamma(b)}{\Gamma(b-a)}\sum_{k=0}^\infty \frac{(a)_k (a-b+1)_k}{k! (-q)^k}\,,
\label{Asymptotic1F1}
\end{align}
where the sign of the complex phase depends  on the argument of $q$. In our case, $q=\ii s$ and the positive sign has to be chosen. In the latter expression, $(x)_k$ is the Pochhammer symbol, which can be defined as $(x)_k \equiv\Gamma(x+k)/\Gamma(x)$.  It follows that at large values of $\rho s=k(r-z)=\rho(1-\cos\theta)$, using (\ref{Asymptotic1F1}) with $q=\ii s\rho$  in (\ref{exact}), one finds at leading order 
\begin{align}
\tilde{\psi}\sim \ &\hbox{e}^{\ii kz + i\gamma\ln \rho s}\left(1-\ii\frac{\gamma^2}{\rho s}\right)\nn\\
&-\frac{\gamma}{\rho s}\frac{\Gamma(1+\ii\gamma)}{\Gamma(1-\ii\gamma)}\hbox{e}^{\ii\rho - \ii\gamma\ln \rho s}\,,
\label{e.exp}
\end{align}
which is valid well outside the region $\rho s=1$ (i.e. it is not valid for $\theta=0$ for finite values of $r$). In (\ref{e.exp}),  one usually identifies  an incident component (first term) and a scattered spherical wave (second term).
The tilde symbol has been used to stress that this is an asymptotic approximate solution, and to distinguish it from the exact one (\ref{exact}). We observe that in Eq.~(\ref{e.exp}) the correction to the incoming wave proportional to $\gamma^2$ can be relevant for large values of $\gamma$, and can be seen as a backreaction of the scattered wave on the incoming wave (when going beyond the Born approximation lowest order). Hence we define the distorted incoming plane wave by
\begin{equation}\label{DefPsiinDisto}
\psi_{{\rm in},\gamma^2} \equiv\hbox{e}^{\ii kz + i\gamma\ln \rho s}\left(1-\ii\frac{\gamma^2}{\rho s}\right)\,.
\end{equation}
In any realistic configuration of Rutherford scattering (and also in the case of  scattering of classical waves, cf. Sec. \ref{class}), we however typically have $\gamma\ll kr$ and this is the regime we will focus on from now on. Hence, in the analytical computations that follow, we will neglect the back-reaction correction  $\propto \gamma^2/(\rho s)$ to the incoming term, assuming that we are working in a regime where this correction is very small.

\begin{figure}[htb]
 	\centering
 	\includegraphics[width=0.45\textwidth]{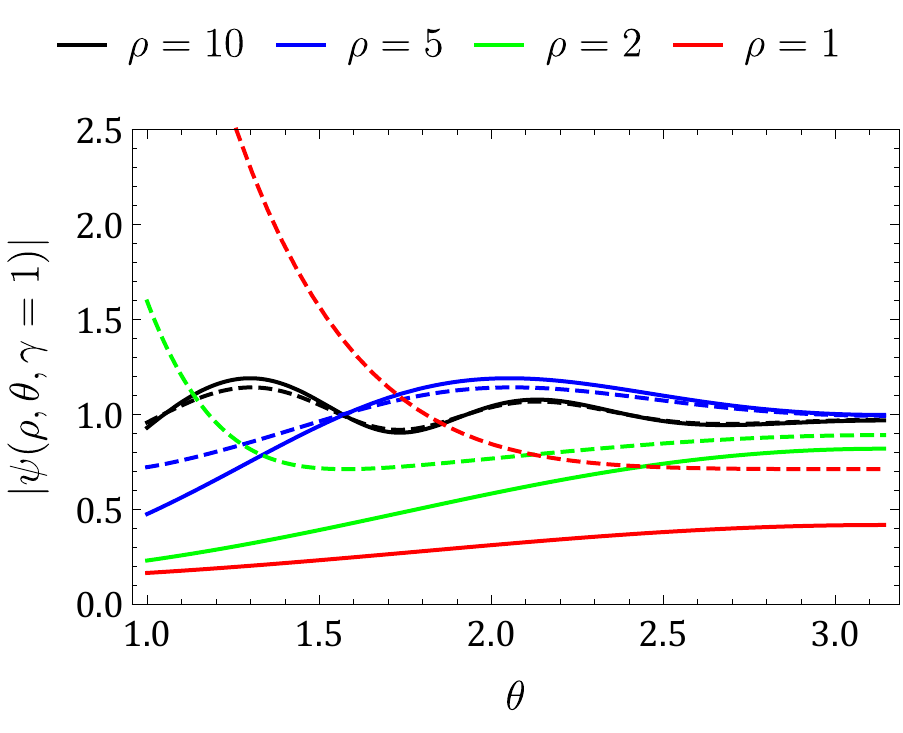}
 	\caption{\label{exactapprox} Comparison between the absolute value of the exact solution~(\ref{e.solgen}) (solid line) and the approximate solution \eqref{e.exp} (dashed line) for $\gamma=1$ and $\rho= 10, 5 , 2, 1$. Numerical parameters values are for illustration purposes, and units arbitrary. Clearly, the approximate solution departs from the exact one at an angle $\theta$ whose value decreases as we increase $\rho$. } 
 	\label{fig:1}
 	\vspace{-0.25cm}
 \end{figure}

Notice that the incoming wave is distorted from a pure plane wave by the presence of the interaction, which gives the appearance of logarithmic phase shifts. 
We notice also that the approximate solution diverges for $\theta\to 0$ (the amplitude of the scattered wave diverges in this limit), unlike the exact solution (\ref{exact}); see Fig.\,\ref{exactapprox}. It clearly follows that the divergence in the cross-section is a direct consequence of using the approximate solution (\ref{e.exp}) outside its regime of validity. 

 Let us see this more explicitly. Using the approximate \emph{Rutherford solution} (\ref{e.exp}) found expanding the exact solution for $\rho s\gg 1$, one identifies the scattering amplitude $f_{\rm R}(\theta)$ as the $r$-independent constant multiplying the (distorted) spherical term
 \be
 \label{e.fFromExact}
 f_{\rm R}(\theta)=\frac{-\,\gamma}{2k\sin^2(\theta/2)}\frac{\Gamma(1+\ii\gamma)}{\Gamma(1-\ii\gamma)}\,,
 \ee
 which in Eq.~(\ref{sigmadef}) gives the usual Rutherford cross-section. However, this derivation shows very explicitly that this result rigorously holds only when $\rho s$ is asymptotically large, which completely excludes the region $\theta =0$.

A priori, one could equivalently have defined the scattering amplitude as
 \be
 \label{e.fFromExact2}
 f(\theta) \equiv \frac{-\,\gamma}{2k\sin^2(\theta/2)}\frac{\Gamma(1+\ii\gamma)}{\Gamma(1-\ii\gamma)} \hbox{e}^{-\ii\gamma \ln(s/2)}\,,
 \ee
 which differs from (\ref{e.fFromExact}) by a phase, which is irrelevant in the computation of the cross-section. This phase redefinition allows one to fully separate the angular part from the radial one. We shall use the definition~\eqref{e.fFromExact2} when comparing the scattering amplitude obtained in real space with the results obtained in multipole space (section~ \ref{s.ellspace}), precisely because the multipole decomposition relies on the separated treatment of radial and angular dependencies.
 
We conclude this section observing  that at leading order in $\rho s$,
  the extra logarithmic phase in Eq.~(\ref{e.exp}) does not affect
  the currents defining the cross-section. To be more precise, if
  $\psi_{\rm in}\equiv\hbox{e}^{\ii kz + \ii\gamma\ln k(r-z)}$, the associated current is given by 
 \be
 \label{JinDistPW}
 {\bf J}_{\rm in}= \frac{\hbar k}{m}\left[\left(\cos\theta +\frac{\gamma}{\rho}\right){\bf e}_r -\left(\sin \theta-\frac{\gamma}{\rho}\frac{\sin\theta}{1-\cos\theta} \right){\bf e}_\theta\right]\,.
 \ee
 This gives the standard $\hbar k/m {\bf e}_z$ at large distance but it has a $1/\rho$ tail so that
 \be
 J_{\rm in}= \frac{\hbar k}{m}\left[1+{\cal O}(1/\rho)\right]\,,
 \ee
 compared to the expression given after Eq.~\eqref{Jscat}.
Similar considerations can be made for the asymptotic outgoing spherical part, allowing us to have an asymptotic cross-section of the form $\propto |f(\theta)|^2$, as expected.


\subsection{Small angle behaviour}

We compare the standard approach in which one takes the large $r$ limit and then faces a singularity in $\theta=0$ to the one in which we look at the solution at small $\theta$ at fixed $r$ but still with $\rho s\ll 1$. 
From the expansion
\beq
 {}_1F_1\left[-\ii\gamma,1;\ii\rho s\right] =1+\gamma\rho s +\frac{\gamma\rho^2}{4}(i+\gamma)s^2+{\cal O}(s^3)\,,
\eeq
the exact solution (\ref{exact}), is approximated when $\rho s \ll 1$ by
\be\label{cone}
\tilde{\psi}=\hbox{e}^{\ii\rho(1-s)}\hbox{e}^{-\pi\gamma/2}\Gamma(1+\ii \gamma)\left(1+\gamma\rho s
\right)\,.
\ee
In the $\theta=0$ direction, we have exactly $s=0$ and the solution reduces to 
\be\label{coneszero}
\tilde{\psi}\rvert_{\theta=0} = \hbox{e}^{\ii k z}\hbox{e}^{-\pi\gamma/2}\Gamma(1+\ii \gamma)\,.
\ee

\subsection{Interpretation of the forward direction}
The large $\rho s$ limit can always be fullfilled for every $\theta\!\in\, ]0, \pi]$. Indeed, in this case we have $s>0$, such that there always exists a distance $r$ large enough to satisfy $\rho s\gg 1$.

We observe that the condition $\rho s = 1$ translates in Cartesian coordinates to $k\big(z+\frac{1}{2k}\big) = \frac{1}{2}k^2(x^2+y^2)$, which defines a paraboloid of revolution. The transverse section of the paraboloid increases as we go towards larger positive $z$, but its angular size, seen from the origin, decreases. The limit $\rho s\gg1$ corresponds to the region sufficiently far outside this paraboloid.

Considering an incoming plane wave along $z$, we follow a point on the wavefront which is identified by a set of coordinates $(x,y)$. 
Such a wavefront point will \emph{enter} the paraboloid once 
\be
k\bigg(z+\frac{1}{2k}\bigg) \geq \frac{1}{2} k^2(x^2+y^2)\,.
\ee
The entrance point thus depends on the point $(x,y)$ that one is following, and is located at a further distance from the scattering center for larger $|x|, |y|$.

\begin{figure}[!htb]
  \includegraphics[width = \columnwidth]{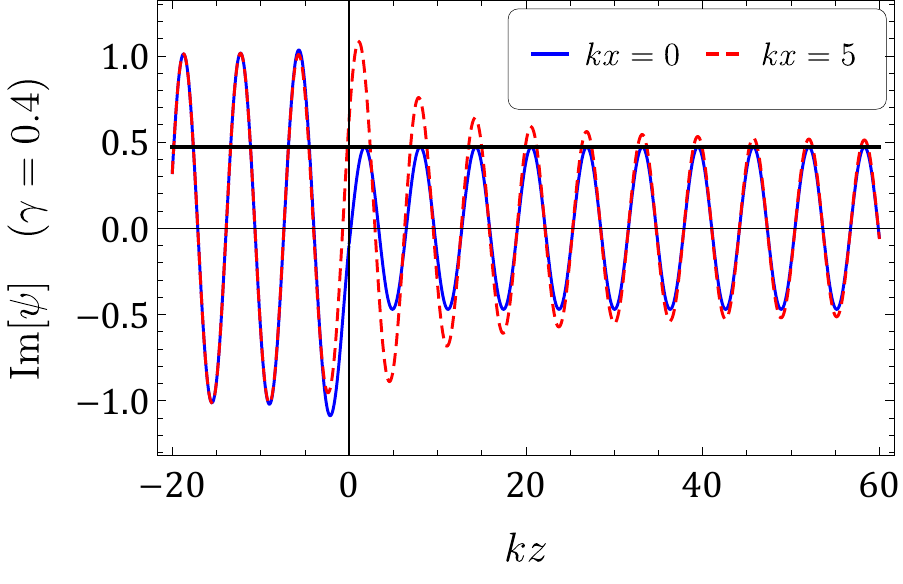}\\
  \includegraphics[width = \columnwidth]{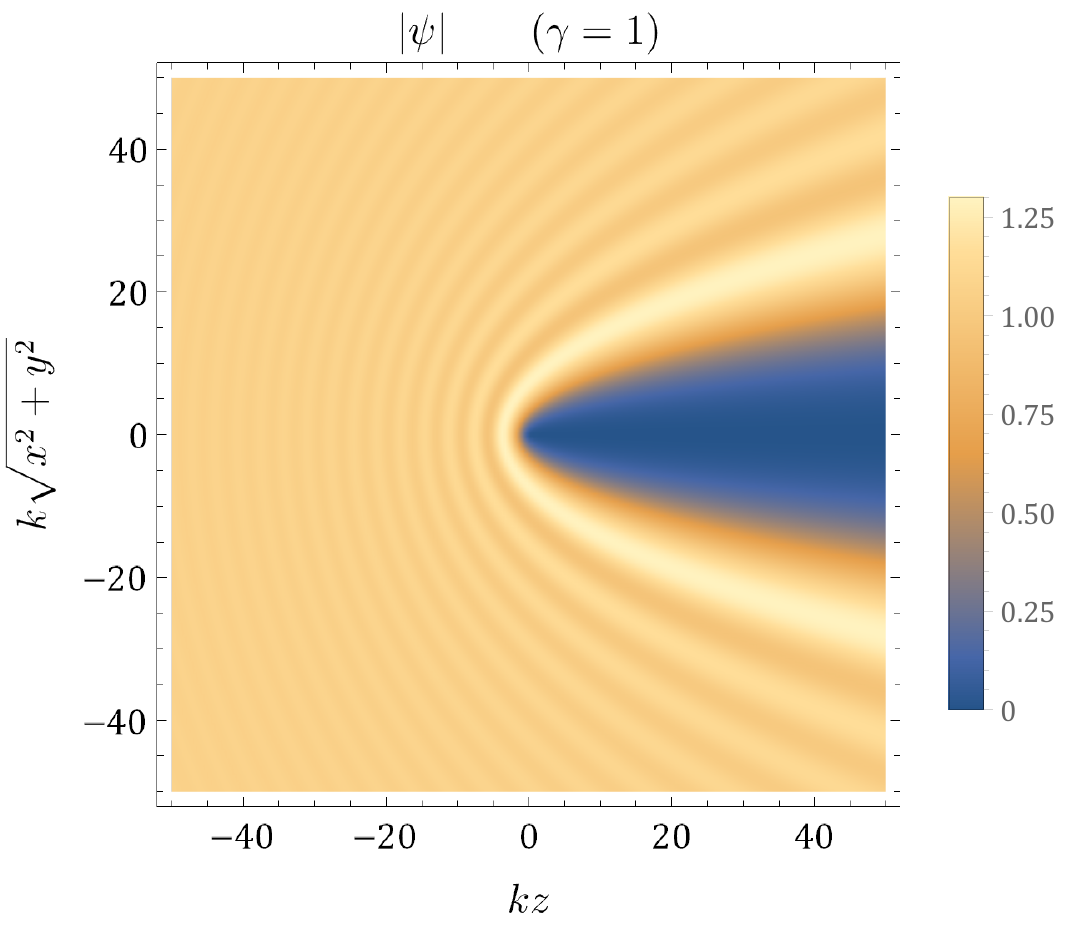}
	\caption{{\it Top:} Exact solution on slices parallel to the $z$ axis $(x = \text{const.}, y=0)$.
		Fixing a higher $kx$, the entrance inside the paraboloid (and thus
		the asymptotic damped plane wave regime) is reached at a higher $z$.
		 The black line is the expected asymptotic amplitude $|\Gamma(1+\ii\gamma)| {\rm e}^{-\pi
			\gamma/2}$. We plot $\text{Im[}\psi\text{]}$, but analogous considerations hold for $\text{Re[}\psi\text{]}$. 
			{\it Bottom:} two-dimensional slice of the exact
                      solution $|\psi|$. The paraboloid region
                      corresponding to $\rho s <1$ is clearly visible.}
		\label{Fig3}
\end{figure}

We illustrate this in Fig.\,\ref{Fig3}. We consider the exact solution \eqref{e.solgen} of the Rutherford problem,  and we look at the evolution of a point in the  $(x, z)$ plane (i.e. we consider $y=0$, to work with slices in the $(x,z)$ plane, the general case follows by azimuthal symmetry).

By construction, $\psi$ has an asymptotic behaviour at $z\to-\infty$ that is oscillating with unit amplitude (see e.g. Eq.~\eqref{e.exp}). On the other hand, sufficiently far inside  the paraboloid, $\psi$ is oscillating with reduced amplitude $\sim \hbox{e}^{-\pi\gamma/2}|\Gamma(1+\ii\gamma)|$ which tends to $0$ for $\gamma \gg 1$, see Eq.~\eqref{coneszero}. 
This corresponds to a damped-wave regime. We can interpret this as the fact that the initial plane wave has transferred a part of its initial magnitude to the scattered wave. The entrance in the paraboloid marks the transition between the two asymptotic regimes.

One can think of the classical analogue of the  wavefront points as a set of initial
particles with an impact parameter $ x = b$. In this picture, part of the initial particles are scattered out in all $\theta \neq 0$ directions, and part of them keep going along the $kx=kb,y=0, kz \to \infty$
direction. The non-scattered part is a fraction $|\Gamma(1+\ii \gamma)| {\rm e}^{-\pi\gamma/2}$ of the initial amplitude. This understanding cannot be reached without the exact solution~\eqref{e.solgen}.

\subsection{Outgoing current and interference} \label{Sect:Inter}

When computing the cross-section, we need to compute the current of the scattered wave. 
Obviously this is not equivalent to the difference between the current of the total wave and the current of the incoming wave. The difference is given by interference terms between the incoming and outgoing wave. 
Explicitly, the current associated to the total wave (approximate solution) $\tilde{\psi}$ can be written as
\be\label{splitcur}
\tilde{\bf J}=\tilde{\bf J}_{\text{in}}+\tilde{\bf J}_{\text{scat}}+\tilde{\bf J}_{\times}\,,
\ee
where $\tilde{\bf J}_{\text{in}}\equiv {\bf J}[\psi_{\text{in}}]$, $\tilde{\bf J}_{\text{scat}} \equiv {\bf J}[\tilde{\psi}_{\text{scat}}]$ and the current operator ${\bf J}$ is defined in Eq.~\eqref{J}. The last term $\tilde{\bf J}_{\times}$ is the current resulting from the interference between the incoming and scattered waves. 

\begin{figure}[ht!]
	\centering
		\includegraphics[width=0.45\textwidth]{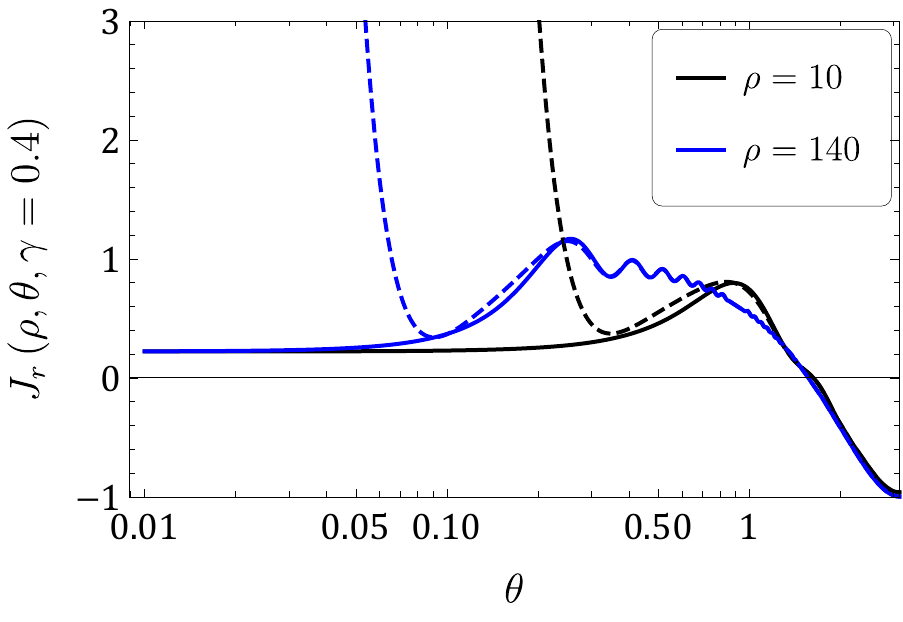}   \\
	\includegraphics[width=0.45\textwidth]{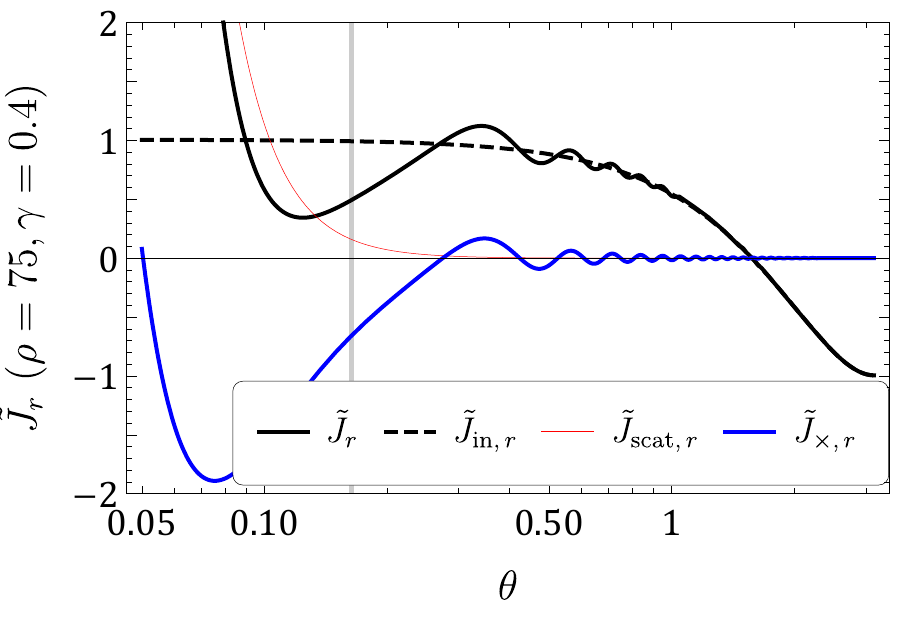}\\
\includegraphics[width=0.45\textwidth]{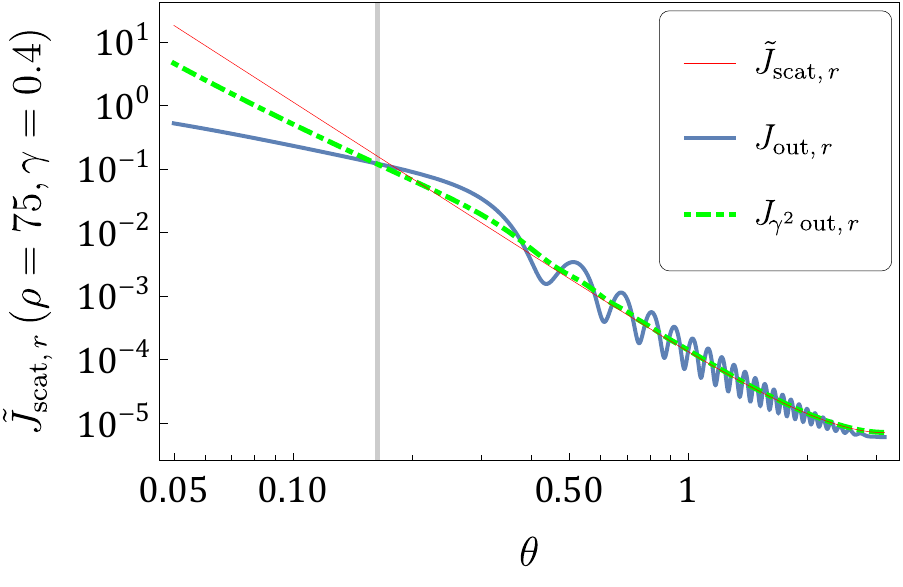}    
	\caption{\label{CurrentComp} \emph{Top}: Comparison between the total currents $ J_r \equiv \bf J [\psi]\cdot\bf e_r $ (solid) and  $\tilde J_r \equiv {\bf J} [\tilde{\psi}]\cdot\bf e_r $ (dashed) for $\rho=$ 10 (black) and 100 (blue). The current computed from $\tilde{\psi}$ (Eq.~\eqref{e.exp}) diverges for small angles, unlike the exact current. \emph{Middle}: Total current computed from $\tilde{\psi}$ (black, solid), along with its different components as in \eqref{splitcur}. The divergence at small angles comes from the divergence of the approximate solution at small angles. 
	\emph{Bottom}: Scattered current from the asymptotic (red) and exact solutions. The latter are computed from $\psi_\text{in}$ without (grey-blue, solid line) and with (green, dash-dotted) backreaction corrections. Vertical bars mark the angle at which $\rho s = 1$ (below which $\tilde{\psi}$ and quantities derived from it are outside their region of validity). For all the plots we chose $\gamma=0.4$. }
\end{figure}

Given an initial waveform $\psi_{\text{in}}$, a scattered current can also be computed from the exact solution (\ref{exact}) subtracting the incoming wave contribution, as 
\be
{\bf{J}}_{\text{out}} \equiv {\bf {J}}[\psi-\psi_{\text{in}}]\,.
\ee
If we chose instead to subtract the distorted plane wave \eqref{DefPsiinDisto}, we define
\be
{\bf{J}}_{\gamma^2\,\text{out}} \equiv {\bf {J}}[\psi-\psi_{\text{in},\gamma^2}]\,.
\ee
In the top panel of Fig.\,\ref{CurrentComp}, we plot the radial component of the total current that can be computed from the exact and asymptotic solutions. The latter diverges as expected at small angles, while the exact one tends to a constant value. The larger $\rho$, the larger is the angular range for which the asymptotic solution is accurate. 
In the middle panel of Fig.\,\ref{CurrentComp} we plot the various contributions to the radial component of the current~(\ref{splitcur}) while in the bottom panel we compare the outgoing current obtained from the exact solution, to the approximated one. 
In the bottom panel of  Fig.\,\ref{CurrentComp}, we compare the radial part of ${\bf{J}}_{\text{out}}$ and ${\bf{J}}_{\gamma^2\,\text{out}}$. 
We note that using the distorted plane wave~\eqref{DefPsiinDisto}, which includes a $\gamma^2$ correction, to define ${\bf{J}}_{\!\gamma^2 \,\text{out}}$ makes the latter closer to ${\tilde{\bf{J}}}_{\,\text{scat}}$, in the sense that it removes a small oscillatory pattern. This pattern is due to the interference between the $\gamma^2$ correction to the incoming wave and the outgoing part in ${\bf{J}}_{ \,\text{out}}$.

As we have seen, interference terms between the incoming wave and the scattered one are neglected in the standard picture, when identifying any observed outgoing flux with the flux of the scattered part of the wave (total minus incident one). The leading order contribution to the interference current  in the radial direction is
\be
\tilde{J}_{\times,r} \simeq -\frac{\hbar \gamma}{m r} \cot^2(\theta/2) \cos[\rho s - 2\gamma \ln (\rho s)+2 \delta_0]\,,
\ee
where $\delta_0$ is a phase shift defined by $\hbox{e}^{2\ii\delta_0} = \frac{\Gamma(1+\ii \gamma)}{\Gamma(1-\ii \gamma)}$. Though $\tilde{J}_{\times,r} $ decays as $1/r$, and thus slower than  $\tilde{J}_{\text{scat},r}$, its angular dependence is oscillatory due to the argument of the last factor.
This oscillatory behaviour is manifest in the middle panel of Fig.~\ref{CurrentComp}. The oscillating length (in the orthoradial direction) is typically
\be
\lambda_{\text{osc}} \simeq \frac{2\pi}{k \sin \theta \left(1- \frac{2\gamma}{\rho s} \right)}\,,
\ee
hence as long as the detector is larger than this length, which is proportional to the wavelength of the scattered wave, this contribution of the current is averaged out. Therefore, interference effects between incoming and outgoing waveforms, are typically not observable experimentally (in quantum scattering experiments). Note that for small angles the oscillating length in the orthoradial direction is much larger than the wavelength of the scattered wave $2\pi/k$. 

There is yet another reason for discarding the interference contribution. Following Ref.~\cite{Schiff}, let us assume that quantum waves generated from a source are collimated by a diaphragm into a rather well-defined beam.
An infinite plane wave $\propto \exp(\ii kz)$ does not describe such a collimated beam. The latter can however be modelled as a superposition of infinite plane waves that have wavevectors of slightly different norms and directions. 

The scattering amplitude $f$ typically has slow angular variations, so that the small directional spread of the incident wavevectors does not affect $f$ strongly. It follows that, at a point of observation outside the beam axis, only the $f$ term matters. On the other hand, if we consider the incident direction far enough from the scattering region, the $f$ term can be made negligible, hence the incident flux can be calculated from the plane-wave term alone.  The bottom line is that in such problems, the interference terms in the region of observation are a consequence of the idealization implicit in assuming an infinite incoming plane wave, and they usually have no observational impact in quantum scattering experiments.

All these considerations do not hold for the case of classical scattering of waves off Coulomb-like potentials. Indeed, even if the mathematical structure of the classical scattering problem is very similar to the quantum counterpart, see Section \ref{class}, the physical setting is very different in the two cases. In classical scattering, a \emph{locally} plane wave diffused by a given potential is a physical (\emph{observable}) object unlike in the quantum scattering case where the fact that the incoming waveform had an infinite extension was an artifact of the mathematical description (the observable object is the flux, collimated).  It follows that interference effects in classical settings will be important in a broad region in front of the target, hence some care has to be taken when introducing a cross-section in this context. Moreover, in such contexts scattering takes place for long wavelength signals and detectors are designed to measure directly the waveform and not the flux, e.g. in radio-astronomy where the electric field is detected or for gravitational waves detection where the metric strain is monitored.

\section{Rutherford in $\ell$ space}
\label{s.ellspace}
\subsection{Formulation of the problem}

The Rutherford cross-section can also be derived in the multipole space, i.e. using an expansion of the angular dependence in Legendre polynomials $P_\ell(\cos\theta)$. A plane wave is expanded as
\begin{align}
\label{planewaveexpansion}
\psi_{\text{plane}}&= \hbox{e}^{\ii kr \cos\theta}=\sum_{\ell}\ii^{\ell}(2\ell+1)j_{\ell}(kr)P_{\ell}(\cos\theta) \,,
\end{align}
which we shorten as $ \sum_\ell \psi_{\text{plane}, \ell}\, P_{\ell}(\cos\theta)$. In scattering problems, one usually wants to identify a spherical incoming and a spherical outgoing wave. For a given $\ell\ll kr $ one can expand the spherical Bessel into the sum of incoming and outgoing spherical waves, and obtains 
\be\label{inell} 
\psi_{\text{plane}, \ell}\simeq
\frac{2\ell+1}{2\ii k r}\ \left[(-1)^{\ell+1} \hbox{e}^{-\ii k r}+\hbox{e}^{\ii k r}\right]\,.
\ee
A similar decomposition can be introduced for a generic wave. For example, the total wave in real space can be written asymptotically as 
\be
\label{e.IdealWaveFunction}
\psi_\text{tot} \simeq \hbox{e}^{\ii kr \cos\theta} +f(\theta)\frac{\hbox{e}^{\ii kr}}{r}\,.
\ee
The decomposition in a basis of Legendre polynomials must be of the form
\be
\label{e.AsymptoticPsiTot}
\psi_{\text{tot}, \ell}\simeq \frac{2\ell+1}{2\ii k r} \left[(-1)^{\ell+1} \hbox{e}^{-\ii k r}+ \hbox{e}^{2\ii\delta_\ell} \hbox{e}^{\ii k r}\right]\,,
\ee
where the functions $\delta_{\ell}$ are called phase shifts. They encode the effects of scattering, which affects by construction only outgoing terms. Phase shifts describe the departure of the wave function from the plane wave, i.e. for $\delta_{\ell}=0$, $\psi_\text{tot}$ coincides with the plane wave. 
The scattering amplitude is traditionally defined as the amplitude of the scattered wave (substracting the transmitted component)
\be\label{ft}
f(\theta)\equiv\sum_\ell \frac{2\ell+1}{2\ii k}( \hbox{e}^{2\ii\delta_\ell}-1) P_{\ell}(\cos\theta)\,. 
\ee

We now discuss how the above procedure can be applied to Rutherford scattering. Let us expand $\psi$ as 
\be\label{WaveSolution}
\psi(\rho,\theta) =\sum_\ell \psi_\ell(\rho)P_{\ell} =\sum_\ell\frac{\Psi_\ell(\rho)}{\rho}P_{\ell}\,,
\ee
where $P_\ell$ are evaluated at $\cos(\theta)$. The differential equation \eqref{e.schro2b} becomes
\be\label{e.schro2bm}
\left[\partial_\rho^2+\frac{2}{\rho}\partial_\rho - \frac{\ell(\ell+1)}{\rho^2} + 1 -2\frac{\gamma}{\rho}\right]\psi_\ell=0\,,
\ee
or equivalently
\be\label{e.schro3m}
\Psi_\ell'' + \left[1 -\frac{2\gamma}{\rho}- \frac{\ell(\ell+1)}{\rho^2}  \right]\Psi_\ell=0\,.
\ee
This is known as the Coulomb wave equation, which is a Whittaker equation whose solutions are known. However rewriting it with a proper rescaling by $\hbox{e}^{-i\rho}\rho^{\ell+1}$, it appears that the solutions are expressed directly in terms of confluent hypergeometric functions as
\be
\label{ExactEllSolution} 
\Psi_\ell = C(\ell, \gamma)\, \rho^{\ell+1} \hbox{e}^{- \ii \rho}\Ff(\ell+1- \ii\gamma, 2\ell+2; 2\ii\rho)\,,
\ee
where $C(\ell, \gamma)$ is at this stage an arbitrary constant.
There exists another independent solution of \eqref{e.schro3m} that we discard since it is not regular as $\rho\to0$. Apart from normalisation conventions, the function $\Psi_\ell$ given by \eqref{ExactEllSolution} is usually called the regular Coulomb wave function in the literature (see e.g. \cite{Gaspard}).

We now consider the asymptotic behaviour \eqref{Asymptotic1F1} for large $\rho$, where large means typically  $\rho \gg \ell(\ell+1) + \gamma^2$ since the expansion~\eqref{Asymptotic1F1} is typically valid when $|(b-a)(1-a)/q| \ll 1$ and $|a (a-b+1)/q| \ll 1$. After some algebra, we get
\be\label{Psiell}
\frac{\Psi_\ell}{\rho}  \sim \frac{\tilde{C}(\ell, \gamma)}{2\ii\rho} \; \left[(-1)^{\ell+1} \hbox{e}^{-\ii  \rho_c}+ \hbox{e}^{2\ii\delta_\ell} \hbox{e}^{\ii \rho_c}\right]\,, 
\ee
with 
\be
\rho_c = \rho - \gamma \ln (2 \rho)\,,
\ee
and where the factors $\hbox{e}^{\ii \delta_\ell}$ are now directly coming from the asymptotic expansion of the $\Ff$ function, and defined such that
\be
\Gamma(\ell+1+\ii\gamma) \equiv |\Gamma(\ell+1+\ii\gamma)| \hbox{e}^{\ii\delta_\ell}\,,
\label{PhaseShift}
\ee
or equivalently
\be
\hbox{e}^{2\ii\delta_\ell} = \frac{\Gamma(\ell+1+\ii\gamma)}{\Gamma(\ell+1-\ii\gamma)}\,.
\label{e.PS}
\ee
The overall constant $\tilde{C}(\ell, \gamma)$ consists of the original $C(\ell, \gamma)$ with additional factors coming from the asymptotic expansion, namely
\be
\tilde{C}(\ell, \gamma) \equiv C(\ell, \gamma)\, \frac{1}{(2\ii)^\ell} \frac{\Gamma(2\ell+2)}{\Gamma(\ell+1+\ii\gamma)} \hbox{e}^{\gamma\pi/2} \,.
\ee
The freedom in $C(\ell, \gamma)$ translates to a freedom in the choice of $\tilde{C}(\ell, \gamma)$. For consistency with our initial conditions, we must require that our solution asymptotically matches Eq.~\eqref{inell} in the limit of the absence of scattering, i.e. for $\gamma = 0$ (implying $\delta_\ell =0$, as well as $\rho=\rho_c$). The result is $\tilde{C}(\ell, \gamma =0) = 2\ell+1$, which can be generalised for $\gamma\neq0$ to 
\be
\tilde{C}(\ell, \gamma) = 2\ell+1\,.
\label{eqCValueScatt}
\ee
This self-consistent choice implies
\be
\label{e.ellSapceRutherfordSolution}
\psi_\ell \sim \frac{2\ell+1}{2\ii k r} \ \left[(-1)^{\ell+1} \hbox{e}^{-\ii \rho_c}+ \hbox{e}^{2\ii\delta_\ell} \hbox{e}^{\ii \rho_c}\right]\,,
\ee
that has the same structure as Eq.~\eqref{e.AsymptoticPsiTot} (the two are identical in the absence of scattering).  For $\gamma\neq0$, the solution of our scattering problem differs from that in Eq.~\eqref{e.AsymptoticPsiTot} by the presence of $\rho_c$ instead of $\rho$ in the exponent. This reflects the fact that the scattering problem does not have solutions in the exact form of a superposition of a plane wave and a spherical one, but instead has logarithmic modifications in the phases with respect to the latter, as we have seen in Eq.~\eqref{e.exp}.

\subsection{Series expression for $f(\theta)$}

Let us consider the series \eqref{ft} for the phase shifts \eqref{e.PS}, namely
\be
f(\theta)=\sum_\ell \frac{2\ell+1}{2\ii k}\bigg(\frac{\Gamma(\ell+1+\ii\gamma)}{\Gamma(\ell+1-\ii\gamma)}-1\bigg) P_\ell(\cos\theta)\,.
\label{nightmare}
\ee
Note that while in \eqref{ft} $f$ is defined as the angular modulation of $\hbox{e}^{\ii\rho}/r$, the Rutherford analogous quantity is, by construction, the angular modulation to $\hbox{e}^{\ii\rho_c}/r$, as can be seen from \eqref{e.ellSapceRutherfordSolution}. The scattered amplitude $f$ resulting from the summation \eqref{nightmare}
should therefore be expected to equal the quantity \eqref{e.fFromExact2}.

Using the results in Appendix \ref{Cyril}, it seems that it is possible to directly resum the series in Eq.~(\ref{nightmare}). Indeed, using Eq.~(\ref{Magic1}) we find 
\bea
&&\frac{1}{(1- \cos \theta)} {\rm e}^{\ii \gamma\ln(2/s)} =\frac{1}{2}\left(\frac{2}{1-\cos \theta}\right)^{1+\ii\gamma}
\\
&&= -\sum_\ell \frac{(2\ell + 1)}{2 \ii \gamma} \frac{\Gamma(\ell+1 + \ii
  \gamma) \Gamma(1- \ii \gamma)}{\Gamma(\ell+1 - \ii \gamma)\Gamma(1+
  \ii \gamma)} P_\ell(\cos \theta)\nn\,.
\eea
Hence we recognise 
\bea
&&-\frac{\gamma}{k(1- \cos \theta)} \frac{\Gamma(1+\ii \gamma)}{\Gamma(1-
  \ii \gamma)} {\rm e}^{\ii \gamma\ln(2/s)} \nn\\
&&= \sum_\ell \frac{(2\ell + 1)}{2 \ii k} \frac{\Gamma(\ell+1 + \ii
  \gamma)}{\Gamma(\ell+1 - \ii \gamma)} P_\ell(\cos \theta) \label{ApplMaster}\,.
\eea
The LHS is nothing else but $f(\theta)$ of Eq.~\eqref{e.fFromExact2}, whereas the RHS is \eqref{ft} up to a term $\propto \sum_\ell (\ell+1/2) P_\ell(\cos \theta) = \delta(1- \cos \theta)$ which vanishes whenever $\theta \neq 0$, and which can be traced to the contribution of the incoming plane wave in the forward direction, which had been subtracted in the definition (\ref{ft}). Therefore this computation seems to indicate that $f(\theta)$ as defined from a series of multipoles in \eqref{ft} agrees with the result \eqref{e.fFromExact2} obtained in real space for $\theta \neq 0$. 

\begin{figure}[!htb]
  \includegraphics[width = \columnwidth]{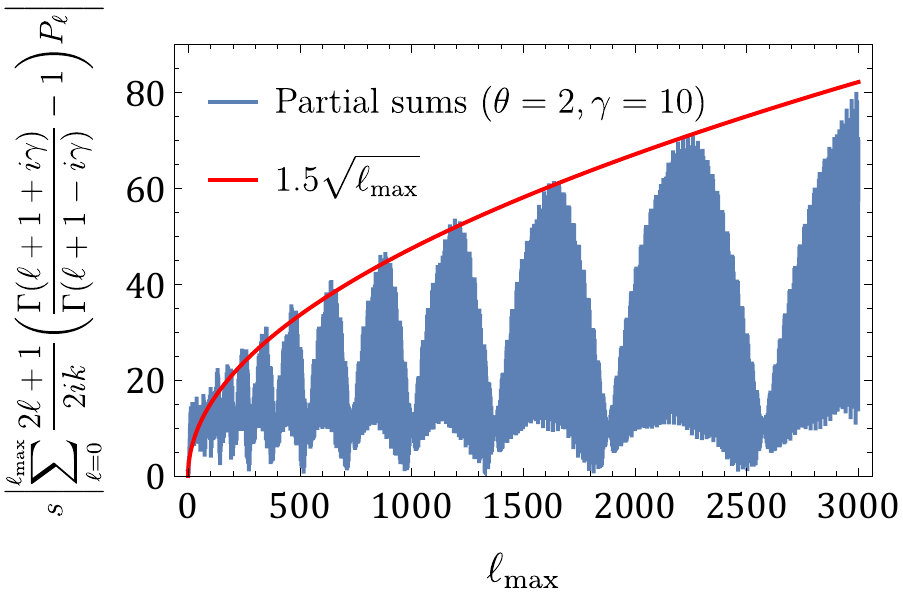}
\caption{Illustration of the non convergence of \eqref{nightmare} in the usual sense of partial sums. We see that the partial sums are highly oscillatory as a function of the cutoff $\ell_\text{max}$. However, the peak of the oscillations grows proportionally to $\sqrt{\ell_\text{max}}$, disabling any convergence for the series in the sense of partial sums. The plot is made at fixed angle $\theta = 2$, and interaction strength $\gamma = 10$, but the overall behaviour is identical as those values are modified. Note the remark in Ref.~\cite{MarquezDivergence} stating that if the partial sums are only summed until peculiar $\ell_\text{max}$ satisfying $\delta_{\ell_\text{max}} = n \pi$ for $n \in\mathbb{N}$, then the diverging oscillatory behaviour is avoided.}
\label{DivergingSum}
\end{figure}

However, in this derivation, formula \eqref{Magic1} was used outside its domain of applicability (obviously the real part of $\ii \gamma$ vanishes). In fact the series \eqref{nightmare} does \emph{not} converge, as was noted e.g. in Refs.~\cite{MarquezDivergence,AhmedCoulombScattering,StrattonSeriesReduction}. Indeed, the large $\ell$ behaviour of Legendre polynomials is
\be\label{AsymptoticPl}
P_\ell(\cos \theta ) \propto \sqrt{\frac{2}{\pi \ell \sin \theta}} \cos \left[\left(\ell+\frac{1}{2}\right)\theta - \frac{\pi}{4}\right],
\ee
and the $\hbox{e}^{2\ii\delta_\ell}$ factors are oscillatory. The $\ell$-th term of series \eqref{nightmare} is thus an oscillatory function with amplitude $\sqrt{\ell}\to\infty$, hence failing to meet a necessary criterion for convergence. Any numerical attempt to sum it with partial sums up to any finite order is therefore doomed to fail, as Fig.~\ref{DivergingSum} illustrates.

This is not surprising, since \eqref{nightmare}  was obtained from an asymptotic expansion valid only for $\ell$ such that $\ell (\ell+1)\ll \rho-\gamma^2$. On the other hand, the series (\ref{WaveSolution}) computed with $\Psi_{\ell}$ given by the exact solution (\ref{ExactEllSolution}) does converge absolutely.
Indeed, for $\ell (\ell+1)+ \gamma^2 \gg \rho$, the $\Ff$ function in \eqref{ExactEllSolution} behaves as $\propto \hbox{e}^{\ii \rho}$ and therefore from the asymptotic expansion $\Gamma(z) \sim \sqrt{2\pi} z^{z-1/2} \hbox{e}^{-z} $, one finds (see e.g. \cite{DLMF_Coulomb})
\begin{equation}
|\psi_\ell| \sim \frac{\hbox{e}^{-\pi\gamma/2}}{\sqrt{2}} \rho^\ell (2\ell)^{-\ell}\hbox{e}^\ell \sim \frac{\rho^\ell \hbox{e}^{-\pi\gamma/2} }{(2\ell-1)!!}\,.
\end{equation}
Given the asymptotic behaviour \eqref{AsymptoticPl}, the absolute convergence of \eqref{WaveSolution} with the exact solutions \eqref{ExactEllSolution} is immediate.

This demonstrates that the $\ell$-space Rutherford scattering problem is not intrinsically ill-defined, and that the series divergence is an artifact, the total underlying series being perfectly summable. Notice that a similar result holds for the plane wave expansion: the series on the right hand side of (\ref{planewaveexpansion})  converges, while its asymptotic version (\ref{inell}), obtained in the limit $\ell\ll kr$ does not.

\begin{figure}[!htb]
  \includegraphics[width = \columnwidth]{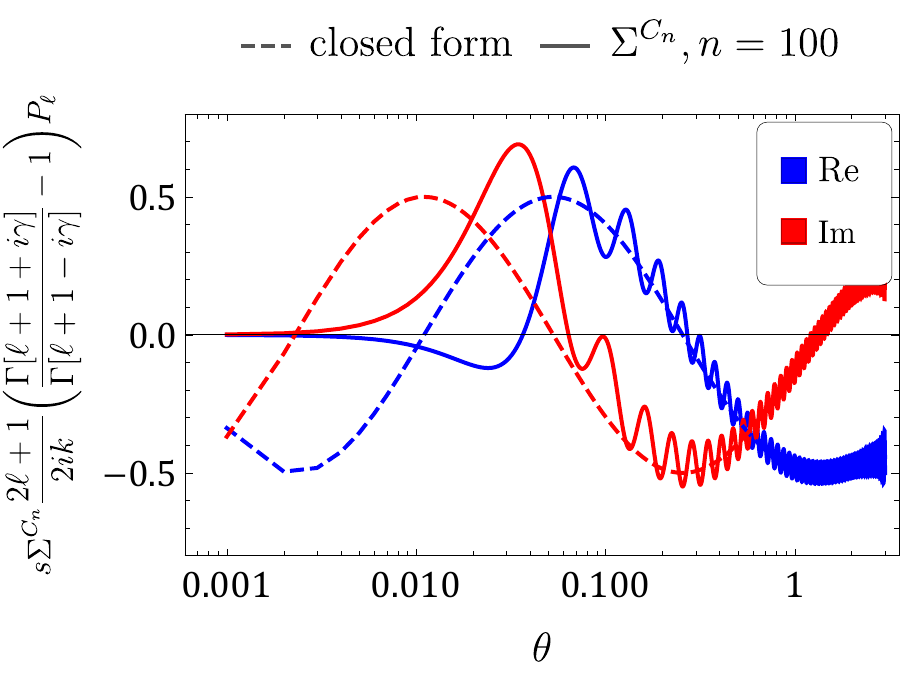}\\
  \includegraphics[width = \columnwidth]{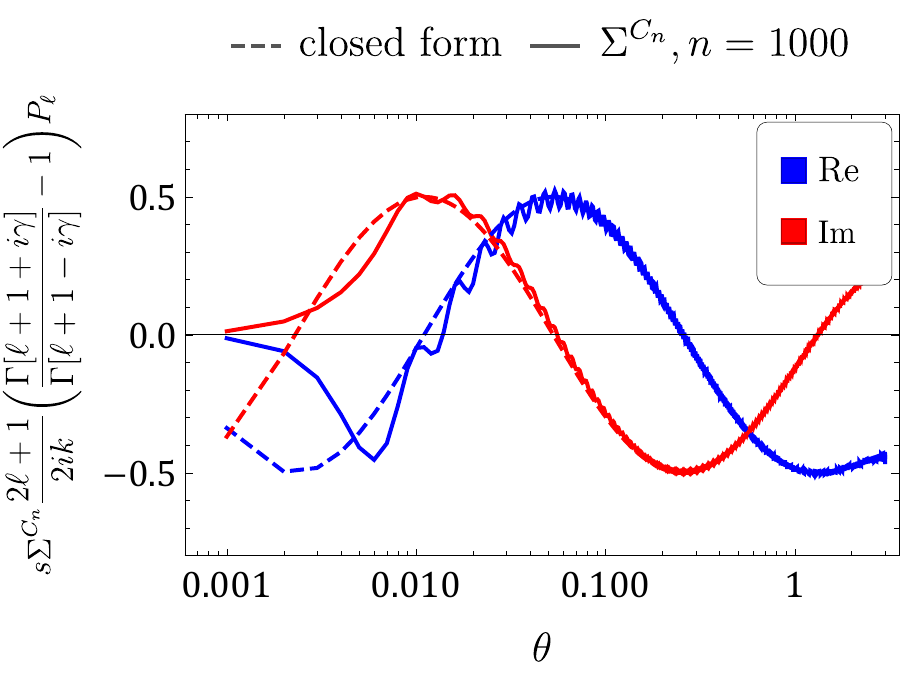}
\caption{Numerical Cesàro partial sums $\Sigma^{C_n}$ (see Appendix \ref{Cesaro}) for the series \eqref{nightmare}, with $n = 100$ (\textit{top}) and $n=1000$ (\textit{bottom}). We explicitly rescaled the result with an extra $s$ overall factor to better appreciate the small angle behaviour. The closed form is then given by \eqref{e.fFromExact2} with an $s$ rescaling.
As $n$ is increased, more multipole terms are taken being summed over, enhancing the angular resolution of the Cesàro partial sums, and thereby improving the agreement with the closed form. Note the logarithmic scale on the angular axis, which puts emphasis on the low $\theta$ end, while a large angular range of the closed form is accurately reproduced by Cesàro summation. These plots were made with $\gamma = 0.5, k = 1$.}\label{FigCesaro}
\end{figure}

Nonetheless, since the incorrect use of \eqref{Magic1} on \eqref{nightmare} lead to the correct result  \eqref{ApplMaster}, this suggests that it should be  possible to use an alternative, non-standard, summation technique that averages out the non-converging oscillations. Cesàro summation, an alternate summation scheme which we review shortly in Appendix \ref{Cesaro}, is designed to average out this type of oscillating behaviour. 
In Fig.~\ref{FigCesaro}, we observe numerically that the Cesàro partial sums converge towards the expected closed form \eqref{e.fFromExact2}, for $\theta \neq 0$, hence showing that the multipole approach is consistent with the real space method of section \eqref{SectionExactToRutherford}.

In order to rely on non-standard summation techniques, one must ensure that the result cannot depend on the scheme chosen. Hereafter we prove that for $\theta\neq 0$, any non-standard summation technique for which the series is summable will converge to the \emph{unique} expression \eqref{e.fFromExact2}, as long as the summation technique satisfies three usual properties. 
Denoting $\widetilde{\sum}_\ell a_\ell$ to be any non-standard summation technique generalising the usual sum $\sum_\ell a_\ell$, these properties are:
\begin{itemize}
    \item Regularity:
    \be
    \widetilde{\sum_\ell}a_\ell = \sum_\ell a_\ell
    \ee
    whenever the latter exists.
    \item Linearity:
    \be
    \widetilde{\sum_\ell}(\lambda a_\ell +b_\ell) = \lambda \widetilde{\sum_\ell}a_\ell+\widetilde{\sum_\ell}b_\ell \,.
    \ee
    \item Stability: 
    \be
     \widetilde{\sum_\ell}a_\ell = a_0 + \widetilde{\sum_{\ell}}a_{\ell+1}\,,
     \ee
     where everywhere the $\ell$ index is assumed to go from 0 to $\infty$.
\end{itemize} 
Cesàro summation is only one example of a non-standard summation scheme satisfying these three properties.

To prove the independence of the result with respect to any convergent, regular, linear and stable summation scheme, let us sum \eqref{nightmare} with such a $\widetilde{\sum}_\ell$ instead of the usual sum. 
First, we know that in real space the scattering amplitude has a singularity for $\theta\rightarrow 0$. Hence in multipole space, each multipole carries some information about this singularity (the spherical harmonic expansion necessarily \emph{runs over} the $\theta=0$ pole). Knowing the expected result allows to formulate the educated guess that the singularity is of the $1/(1-\cos \theta)$ type. Hence by linearity we multiply \eqref{nightmare} by $(1- \cos \theta)$ on every term, at the price of introducing an overall $1/(1-\cos \theta)$ in front of the sum. This can hold only for $\theta\neq0$. We then make use of Bonnet's recursion formula 
\be
\cos \theta P_\ell = \frac{\ell}{2 \ell+1} P_{\ell - 1} + \frac{\ell +
1}{2 \ell + 1} P_{\ell+1}\,,
\ee
to split the sum into three parts. Finally, we use the phase shifts recurrence relations 
\begin{align}
&{\rm e}^{2 \ii \delta_{\ell+1}} = {\rm e}^{2 \ii \delta_\ell}\left(
\frac{\ell+1 + \ii \gamma}{\ell+1 - \ii \gamma}\right)\,,\\
&{\rm e}^{2 \ii \delta_{\ell-1}} = {\rm e}^{2 \ii \delta_\ell}\left(
\frac{\ell - \ii \gamma}{\ell +\ii \gamma}\right)\,,
\end{align}
with the linearity and stability properties of $\widetilde{\sum}_\ell$ to get
\be
(1- \cos \theta) f = \frac{\gamma}{k} \widetilde{\sum_{\ell}}  {\rm e}^{2 \ii
  \delta_\ell}\left(\frac{\ell}{\ell+ \ii \gamma} -
  \frac{\ell+1}{\ell+1 - \ii \gamma}\right)  P_\ell(\cos \theta)\,,
  \label{series}
\ee
This series is convergent in the usual sense of summation, so the same must hold for the $\widetilde{\sum}$ by regularity. 

Indeed, from the identity
\bea
&&\frac{\Gamma(\ell+1+\ii\gamma)}{\Gamma(\ell+1-\ii\gamma)}\bigg(\frac{\ell}{\ell+\ii\gamma}-\frac{\ell+1}{\ell+1-\ii\gamma}\bigg)  \\
&&= -(2\ell+1)\frac{\Gamma(1+\ii\gamma)}{\Gamma(1-\ii\gamma)}\frac{\Gamma(\ell+\ii\gamma)\Gamma(1-\ii \gamma)}{\Gamma(\ell+2-\ii\gamma)\Gamma(\ii \gamma)}\,,\nonumber
\eea
and the tabulated sum~\eqref{inter}, we get the closed form of \eqref{series}:
\be
(1-\cos\theta)f(\theta) = -\frac{\gamma}{k}\frac{\Gamma(1+\ii\gamma)}{\Gamma(1-\ii\gamma)} \hbox{e}^{-\ii\gamma\ln(s/2)}\,,
\label{summedseries}
\ee
that is we prove that all regular, linear and stable non-standard summation schemes, if they converge, must agree and allow one to obtain the expression~\eqref{e.fFromExact2}.

\begin{figure}[!htb]
  \includegraphics[width = \columnwidth]{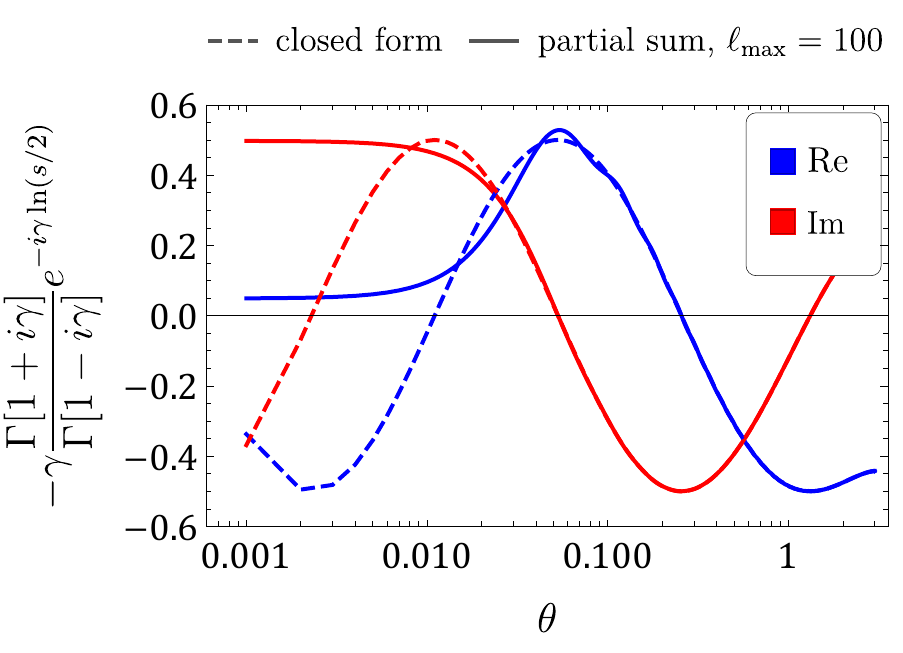}\\
  \includegraphics[width = \columnwidth]{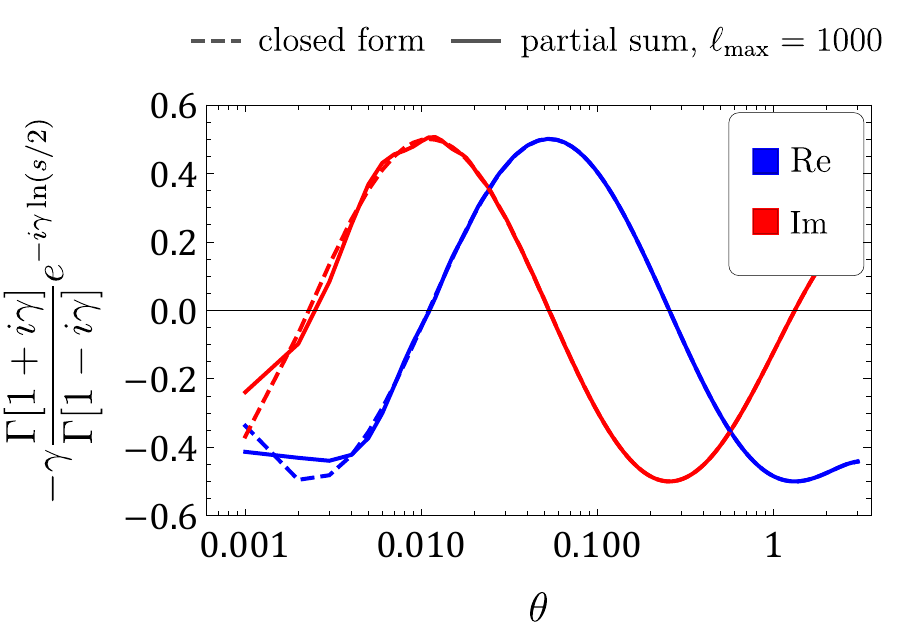}
\caption{Reconstruction of the scattering signal from the partial sums over
  $\ell$. \textit{Top }: $\ell_{\rm max} = 100$. \textit{Bottom} : $\ell_{\rm max} =
  1000$. Blue is real part and red is imaginary part. The partial sums (in the usual sense) corresponding to $(1-\cos\theta)f$ on the RHS of \eqref{series} are in solid lines, while in dashed is the corresponding closed form, i.e. the RHS of \eqref{summedseries}.  We used $\gamma = 0.5$. }\label{FigL}
\end{figure}

 The convergence (in the sense of the usual summation) of series \eqref{series} is illustrated in Fig.\,\ref{FigL}, where it is (numerically) summed up to different values of $\ell_{\text{max}}$. Increasing the value of $\ell_{\text{max}}$ allows one to improve the angular resolution, hence the agreement at small angular scales.

Note that linearity and stability were crucial in the previous steps, and these properties cannot be used if the sum is the usual sum, since the latter lacks convergence. 

This proof largely builds up on considerations made by Refs.~\cite{Yennie, LinCoulombScattering,AhmedCoulombScattering, StrattonSeriesReduction}. However, rather than using the usual sum with a set of mathematically not allowed but convenient tricks, we showed that relying on the more rigorous framework of non-standard summation methods, the result is brought to a unicity statement (under aforementioned assumptions). Turning to non-standard summation methods is made necessary to compensate the fact of having improperly taken asymptotic limits. 

Let us stress again that there is fundamentally no divergence in the exact series [the series \eqref{WaveSolution} with terms \eqref{ExactEllSolution}], and that these convergence issues arise only because of we first used an asymptotic form for the terms of the series which is valid for $\rho \gg \ell(\ell+1) + \gamma^2$, but we then insisted in taking the limit $\ell \to \infty$ when evaluating the sum of approximated terms, that is \eqref{nightmare}.

Finally, let us stress that the summability of the series in the sense of Abel summation, another non-standard summation scheme, was studied in Ref.~\cite{GesztesyAbelSummability}.

\section{Classical wave scattering}\label{class}

Problems of scattering of classical waves off matter overdensities  have mathematically the same structure as the quantum Rutherford scattering, see e.g. \cite{futterman_handler_matzner_1988} for a review on scattering of waves of different spins (see also for recent works on spin-2 fields \cite{PhysRevD.77.044004, Dolan:2008kf}). If follows that several mathematical techniques developed in the framework of quantum scattering can be imported in this context, with a few caveats.

Let us consider the simple case of classical scattering of a massless scalar wave off a black hole. The process is described by the following equation
\be
  \bar{\Box}\Phi=0\,,
  \ee
  where the d'Alembertian is defined on black hole background and we are using Schwarzschild coordinates
  \be
  ds^2=\bar{g}_{\mu\nu}\dd x^{\mu}\dd x^{\nu}=-A(r)\dd t^2+A^{-1}(r)
  \dd r^2 +r^2 \dd\Omega^2\,,
\ee
with $A(r)=1-r_s/r$ with $r_s\equiv2MG$ where $M$ is the black hole mass and $G$ the Newton constant. 
  The problem is static and spherically symmetric, and we assume to have a monochromatic wave with frequency $\omega$. As in section \ref{s.ReviewRutherford}, we can choose our coordinate system with the polar axis aligned with the propagation direction of the incoming wave. With this choice, the problem depends only on the polar angle $\theta$, hence we can introduce the decomposition
  \be
  \Phi=\sum_{\ell}P_{\ell}(\cos\theta)\frac{\hbox{e}^{-\ii\omega t}}{r}u_{\ell}(r, \omega)\,.
  \ee
  We will omit indices on $u$ for simplicity. It can be verified that \cite{Matzner} 
  \be
  \frac{\dd^2 u}{\dd r_*^2}+(\omega^2-V_{\text{eff}})u=0\,,
  \ee
  where $r_*=r+r_s\ln(r/r_s-1)$ and 
  \be
  V_{\text{eff}}=\frac{1}{r^2}\left(1-\frac{r_s}{r}\right)\left(\frac{r_s}{r}+\ell(\ell+1)\right)\,.
 \ee
 If we use the replacement 
 \be
 \label{ubar}
 u=\frac{\bar{u}}{\left(1-r_s/r\right)^{1/2}}\,,
 \ee
 then we get in the limit $r\rightarrow \infty$
 \be
 \label{ubarDiffEq}
 \frac{\dd^2\bar{u}}{\dd r^2}+\left[\omega^2+\frac{4M \omega^2}{r}+\frac{12 M^2 \omega^2}{r^2}-\frac{\ell(\ell+1)}{r^2}+\mathcal{O}(r^{-3})\right]\bar{u}=0\,.
 \ee
In the long wavelength limit i.e. $\ell(\ell+1)>12 (M\omega)^2$ we can neglect the third term in square bracket and we are left with (for $\ell\neq 0$) 
 \be
 \frac{\dd^2\bar{u}}{\dd r^2}+\left[\omega^2+\frac{4M \omega^2}{r}-\frac{\ell(\ell+1)}{r^2}\right]\bar{u}=0\,,
 \ee
which has the same structure as the equation describing Rutherford scattering, see Eq.~(\ref{e.schro3m}). 

Let us use the notation $\gamma = - 2M\omega$, and rewrite our radial equation in terms of the rescaled variable $\rho \equiv \omega r$. We obtain
\be
\frac{\dd^2\bar{u}}{\dd\rho^2}+\left[1-\frac{2\gamma}{\rho}-\frac{\ell(\ell+1)}{\rho^2}\right]\bar{u}=0\,.
\label{CoulombEqRho}
\ee
Then all results derived for Rutherford hold in this case. In particular, if the incoming wave is a (distorted) plane wave, the resulting asymptotic solution will be of the form 
\be
\frac{u}{r} = \frac{2\ell+1}{2\ii\omega r} \left[(-1)^{\ell+1} \hbox{e}^{-\ii \omega r_c}+ \hbox{e}^{2i\delta_\ell} \hbox{e}^{\ii \omega r_c}\right]\,,
\label{CompactSolution}
\ee
where $r_c\equiv (\rho-\gamma\ln(2\rho))/\omega$ with phase shift
\be
\hbox{e}^{2\ii\delta_\ell} = \frac{\Gamma(\ell+1+\ii\gamma)}{\Gamma(\ell+1-\ii\gamma)}\,.
\ee

In the literature, one often pushes this analogy with a quantum scattering problem forward, and introduces a differential cross-section defined as the ratio of the outgoing and incoming flux. While this is a perfectly well-defined object from a mathematical point of view, it does not correspond to any physical observable. When considering scattering of classical waves, the quantity that we measure is often the wave itself rather than a cross-section. Moreover, as we have seen, when introducing a cross-section we implicitly neglect the effect of interference between the incoming and outgoing wave. While this is well-justified in the context of a quantum scattering experiment, in the context of scattering of classical wave there is no reason to assume that interference is small because the detector is typically smaller than the wavelength. On the contrary, the wave observed coming out of the scattering will be the superposition of a transmitted wave and a diffused one, and the two will interfere in an extended region after the scattering center.

\section{Discussion and conclusions}

In this paper, we have reviewed with a pedagogical approach the problem of quantum scattering off a Coulomb-like potential. We have shown that in real space it is possible to find an exact solution and we identified the regime of validity of the so-called \emph{Rutherford solution}, used in the computation of the scattering cross-section. We have stressed that the divergence of the scattering cross-section in the forward direction is a consequence of extrapolating the Rutherford solution outside its regime of validity. As a consequence, there is no need to invoke physical mechanisms to explain such a divergence, which is a pure mathematical artifact. 
We have also reviewed the computation of Rutherford cross-section in multipole space, providing a detailed explanation of how the sum over $\ell$ can be resummed to recover the result for the scattering amplitude known in real space. 

Finally, we have commented on the role of interference and explained that, while in a quantum context interference effects can be legitimately neglected, this is not the case in the context of classical scattering of waves off mass singularities. This leads to some important differences between the problems of quantum and classical scattering. Indeed, while from a formal point of view, the problem of classical scattering is very similar to a quantum scattering in $\ell$-space, the physical observable quantities in the two contexts are not the same. For a Rutherford-like scattering it is a flux (incoming and outgoing) the quantity that we observe. Interference can be neglected since in any observational setting the incoming beam is typically sufficiently well collimated. As a consequence, the notion of a  cross-section can be introduced and it corresponds to an observable quantity. On the other hand, in a classical context, what we observe is  a waveform and in a broad region after the target we observe a superposition of transmitted and scattered waves. 
In this context the notion of a cross-section does not correspond to any physical observable:  while of course one can formally introduce  and compute it, some care has to be taken when assigning it a physical meaning and when using it to describe measurable effects. \\

\noindent{\bf{Acknowledgements}} We thank Ruth Durrer, Djibril Ben Achour and Guillaume Faye for discussions.  M.P. and G.C. acknowledge support from the Swiss National Science Foundation (Ambizione grant, \emph{Gravitational wave propagation in the clustered universe}). 

\appendix

\section{Derivation of the exact Rutherford scattering solution}
\label{ExactSolDerivation}

We summarize the derivation of the exact analytical solution to the Rutherford differential equation \eqref{e.schro2}, which we recall here 
\be
\label{StartingEquation}
\left(\Delta + k^2  - 2\frac{\gamma k}{r}\right)\psi_{\bf k} = 0\,,
\ee
where the ${\bf k}$ subscript is added to the function $\psi$ to highlight that we seek for a solution at given ${\bf k}$ (i.e. given energy). A general solution is a superposition of $\psi_{\bf k}$ solutions.

Let us use parabolic coordinates, defined by
\be 
\xi\equiv r\big(1+\cos\theta\big)\,,\quad \zeta\equiv r\big(1-\cos\theta\big)\,,\quad \varphi\,.
\ee
In these coordinates, the Laplacian operator reads
\be
\Delta = \frac{4}{\xi+\zeta}\bigg[\frac{\partial}{\partial \xi}\bigg(\xi \frac{\partial \,}{\partial \xi}\bigg)+ \frac{\partial}{\partial \zeta}\bigg(\zeta \frac{\partial \,}{\partial \zeta}\bigg)\bigg]+\frac{1}{\xi\zeta}\frac{\partial^2}{\partial \varphi^2}\,.
\ee
We can set our coordinate axes such that for given ${\bf k}$, $\varphi$ is the azimuthal angle with respect to the direction of ${\bf k}$. Then, by symmetry, a solution should be independent of $\varphi$, i.e. one should have $\psi_{\bf k}(\xi, \zeta)$.

One can solve this equation by separation of variables and seek for solutions $\psi_{\bf k}(\xi, \zeta) = X(\xi)Z(\zeta)$ for some functions $X, Z$ to be determined (we omit an explicit ${\bf k}$ subscript for notation simplicity). 
We can then rewrite \eqref{StartingEquation} as
\begin{align}
&\left\{\left[\frac{\partial}{\partial \xi}\left(\xi \frac{\partial}{\partial \xi}\right)+ \frac{\partial}{\partial \zeta}\left(\zeta \frac{\partial}{\partial \zeta}\right)\right] +\frac{k^2}{4}(\xi+\zeta)-k\gamma\right\}X(\xi)Z(\zeta)\nn\\
&=0\,,
 \end{align}
 which gives
\begin{align}
&0=Z(\zeta)\ \Bigl\{ \frac{\partial}{\partial \xi}\bigg(\xi \frac{\partial X}{\partial \xi}\bigg) +\frac{k^2}{4}\xi X-\alpha X\Bigr\}+ \nn\\
+ & \ X(\xi)\ \Bigl\{\frac{\partial}{\partial \zeta}\bigg(\zeta \frac{\partial Z}{\partial \zeta}\bigg) +\frac{k^2}{4}\zeta Z-(k\gamma -\alpha) Z\Bigr\}\,,
\end{align}
where $\alpha$ can be any arbitrary number with the dimension of $k$ ($\alpha$ is not a parameter of the problem, and the full equation is independent of it, only the splitting into the two above parts is). The reason for introducing this $\alpha$ is that it allows one to separate variables, namely obtaining two independent differential equations of one variable. Indeed, one can then restrict further the class of solutions we look for by demanding that both quantities into brackets vanish, which leads to one equation for $X(\xi)$, and one for $Z(\zeta)$.

Performing the field redefinition  $X(\xi)\equiv  \hbox{e}^{-ik\xi/2}\tilde{X}(\xi), \,Z(\zeta)\equiv  \hbox{e}^{-ik\zeta/2}\tilde{Z}(\zeta)$, the equations for $\tilde{X}, \tilde{Z}$ become
\bea
\xi\frac{\partial^2 \Tilde{X}}{\partial \xi^2}+(1-ik\xi)\frac{\partial \Tilde{X}}{\partial \xi}-\bigg(\frac{i k}{2}+\alpha\bigg)\Tilde{X}  = 0\,,\\
\zeta\frac{\partial^2 \Tilde{Z}}{\partial \zeta^2}+(1-ik\zeta)\frac{\partial \Tilde{Z}}{\partial \zeta}-\bigg(\frac{i k}{2}+(k\gamma-\alpha)\bigg)\Tilde{Z}  = 0\,.
\eea

The latter are confluent hypergeometric equations of the generic form 
\be
\label{ComplexCHE}
zF''(z)+(\tilde{B}-z)F'(z)- \tilde{A}F(z)=0 \,,
\ee
for some generically complex coefficients $\tilde{A}, \tilde{B}$ (take $z= ik\xi, \ z=ik\zeta$ respectively). Note that in particular, the parameters $\tilde{A}, \tilde{B}$ $\notin \mathbb{Z}_-$ for our physical problem, and it is therefore well defined to take expressions like $\Gamma(\tilde{A}), \Gamma(\tilde{B})$.
The confluent hypergeometric function of the first kind
\be
F(z)=\Ff(\tilde{A}, \tilde{B}; z)\equiv \sum_{n=0}^\infty \frac{\Gamma(\tilde{A}+n)\Gamma(\tilde{B})}{\Gamma(\tilde{A})\Gamma(\tilde{B}+n)}\;\frac{z^n}{n!}\,,
\ee
is a solution of the confluent hypergeometric equation for such complex parameters $\tilde{A}, \tilde{B}$, and is regular at the origin (as can be checked with an explicit substitution).

We thus end up obtaining
\bea
X(\xi) = C_1 \,\hbox{e}^{\frac{-ik\xi}{2}}\,\Ff\bigg(\frac{1}{2}-\frac{i\alpha}{k},1; ik\xi\bigg)\,,\\
Z(\zeta) = C_2 \,\hbox{e}^{\frac{-ik\zeta}{2}}\,\Ff\bigg(\frac{1}{2}-i\gamma+\frac{i\alpha}{k},1; ik\zeta\bigg)\,,
\eea
for two arbitrary multiplicative constants $C_{1,2}$, allowing thereby to find $\psi_{\bf k}$ for ${\bf k}$ along $\hat{z}$.
Replacing spherical coordinates, and making the systematic replacement $k r \cos(\theta) \to {\bf k}\cdot{\bf r}$, the full solution for arbitrary ${\bf k}$ then becomes
\be\label{app:fin}
\begin{split}
    \psi_{\bf k} = C({\bf k}) \hbox{e}^{-ikr} &\Ff(\beta({\bf k}), 1; i (kr+{\bf k}\cdot{\bf r}))\\
    &\Ff(1-\beta({\bf k})-i\gamma, 1; i (kr-{\bf k}\cdot{\bf r})) \,,
\end{split}
\ee
where in full generality the arbitrary constants $C = C_1 C_2$ and $\alpha$ can be given a ${\bf k}$ dependency, and we replaced  $\alpha({\bf k})$ by the equally arbitrary parameter $\beta({\bf k}) = \frac{1}{2}-i\frac{\alpha}{k}$.

If some conditions on the behaviour of the solution are given (e.g. typically one can have asymptotic requirements), these can be used to constrain the arbitrary (but dimensionless) functions $C, \beta$. A very generic constraint is given by the fact that at least in the absence of interaction ($\gamma=0$), the solution should be bounded asymptotically. From the asymptotic form of $\Ff$, one can verify that this implies $\text{Re}(\beta)\in [0,1]$.

As a further example, requiring $\psi_{\bf k}$ to be a plane wave $C({\bf k}) \hbox{e}^{i\bf k\cdot r}$ in the absence of scattering ($\gamma = 0$) imposes to have strictly $\beta({\bf k}) = 1$. This can be checked considering the behaviour of those functions in the direction ${\bf k}\cdot{\bf r} =  kr$, as well as from a comparison of the series definition of $\Ff$ with that of the exponential.

Extending this $\beta$ for the full $\psi_{\bf k} $ with $\gamma$ generic, we have $\Ff(1, 1, z) = \hbox{e}^z$ in Eq.~(\ref{app:fin}), hence 
\be
\psi_{\bf k} \propto \hbox{e}^{i\bf k\cdot r} \ \Ff(-i\gamma, 1; i (kr-{\bf k}\cdot{\bf r}))\,.
\ee
This is the solution reported in section \ref{SectionExactToRutherford}. The asymptotic behaviour \eqref{e.exp} allows for an immediate comparison with the approximate results of section \ref{s.ReviewRutherford}.

\section{Resummation of Legendre-dependent functions}\label{Cyril}

In this appendix we list the key formulas used in the body of the paper to resum the Rutherford amplitude in multipole space. 
The key result is the identity (for $\text{Re}(a)>0$)
\be
(1-\mu)^{a-1} = \sum_\ell  c_\ell P_\ell(\mu)\,.\label{Magic1}
\ee
with
\be\label{cl}
c_\ell = 2^{a-1}(2 \ell+1)
\frac{\Gamma(a)\Gamma(\ell+1-a)}{\Gamma(1-a)\Gamma(\ell+1+a)}\,.
\ee
These coefficients were extracted with the following integral (see 7.127 of \cite{gradshteyn2007}) valid for $\text{Re}(a)>0$
\be
c_\ell \equiv \frac{(2\ell+1)}{2}\int_{-1}^1 \left(1-x\right)^{a-1} P_\ell(x)\dd x\,.
\ee
A direct proof evaluation of these integrals consists in using the Rodrigues formula 
\be
P_\ell(x) = \frac{1}{2^\ell \ell!} \frac{\dd ^\ell}{\dd x^\ell}[(x^2-1)^\ell]\,,
\ee
and subsequently integrating by parts $\ell$ times. All boundary terms vanish, hence we find 
\be
c_\ell = \int_{-1}^1
\frac{(2\ell+1)\Gamma(\ell+1-a)}{2^{\ell+1} \Gamma(\ell+1)\Gamma(1-a)} (1+x)^\ell (1-x)^{a-1}\,.
\ee
With the change of variables $z = (1+x)/2$ this leads to an Euler integral $B(x,y) \equiv \int_0^1 z^{x-1}
(1-z)^{y-1} \dd z  = \Gamma(x)\Gamma(y)/\Gamma(x+y)$ and the result \eqref{cl} follows. 

From the master formula \eqref{Magic1} we find in particular 
\bea\label{inter}
{\rm e}^{\ii \gamma \ln(2/s)} &=& \left(\frac{1- \cos \theta}{2} \right
)^{-\ii \gamma } \\
&=&- \ii \gamma \sum_\ell (2\ell+1) \frac{\Gamma(\ell+\ii \gamma )\Gamma(-\ii \gamma)}{\Gamma(\ell+2-\ii \gamma)\Gamma(\ii \gamma)}P_\ell(\cos \theta)\,.\nonumber
\eea

\section{Cesàro summation}\label{Cesaro}

In section \ref{s.ellspace}, we mentioned Cesàro summation as a summation technique that can average out the oscillations that are preventing some series from converging.
We briefly introduce it here. Further details can be found in the classic reference \cite{Hardy}.

Cesàro summation is a regular, linear and stable summation method defined as follows.
Suppose one has a series in $\ell$ with terms $a_\ell$. While the usual summation considers the partial sums $\sigma_k = \sum_{\ell = 0}^k a_\ell$, and takes $k\to\infty$, Cesàro summation considers
\be
\label{CesaroPartialSum}
\frac{1}{n+1} \sum_{k=0}^{n}\sigma_k =\frac{1}{n+1} \sum_{k=0}^{n}\sum_{\ell = 0}^k a_\ell \,,
\ee
and takes $n\to\infty$. By construction, Cesàro summation is an average of the ordinary partial sums.
Here we call the result of \eqref{CesaroPartialSum} with a \emph{finite} value for $n\in\mathbb{N}$ the $n$-th Cesàro partial sum and use the shorthand $\Sigma^{C_n}a_\ell$ to denote it. 

Turning to our series of interest \eqref{nightmare}, and substituting the ordinary partial sums with Cesàro partial sums, we numerically see that the series becomes convergent for $\theta\neq0$. The Cesàro averaging process averages out the ever growing oscillations of the partial sums. The results of the numerical summation are illustrated in Fig.~\ref{FigCesaro}.

\bibliographystyle{bib-style}
\bibliography{myrefs}
\end{document}